\documentclass[preprint,prl,groupedaddress,showpacs]{revtex4}
\usepackage{amssymb}
\usepackage{amsmath}
\usepackage{graphicx}
\usepackage{dcolumn}
\usepackage{color}

\begin{document}

\title{Robust Majorana signature detection with a coupled quantum
dot-nanomechanical resonator in all-optical domain}
\author{$^{1}$Hua-Jun Chen$^{\dagger }$, $^{1}$Chang-Zhao Chen, $^{1}$%
Xian-Wen Fang, $^{1}$Yang Li, $^{1}$Guang-Hong Miao, $^{2}$Shao-Fei Tu, and $%
^{2}$Ka-Di Zhu}
\email{chenphysics@126.com}
\affiliation{$^{1}$School of Science, Anhui University of Science and Technology, Huainan
Anhui, 232001, China }
\affiliation{$^{2}$Key Laboratory of Artificial Structures and Quantum Control (Ministry
of Education), Department of Physics and Astronomy, Shanghai Jiao Tong
University, 800 DongChuan Road, Shanghai 200240, China }

\begin{abstract}
Motivated by a recent experiment [Nadj-Perge et al., Science 346, 602
(2014)] providing evidence for Majorana zero modes in iron chains on the
superconducting Pb surface, in the present work, we theoretically propose an
all-optical scheme to detect Majorana fermions, which is very different from
the current tunneling measurement based on electrical means. The optical
detection proposal consists of a quantum dot embedded in a nanomechanical
resonator with optical pump-probe technology. With the optical means, the
signal in the coherent optical spectrum presents a distinct signature for
the existence of Majorana fermions in the end of iron chains. Further, the
vibration of the nanomechanical resonator behaving as a phonon cavity will
enhance the exciton resonance spectrum, which makes the Majorana fermions
more sensitive to be detectable. This optical scheme affords a potential
supplement for detection of Majorana fermions and supports to use Majorana
fermions in Fe chains as qubits for potential applications in quantum
computing devices.
\end{abstract}

\maketitle




\section{INTRODUCTION}

Majorana fermions (MFs) are real solutions of the Dirac equation and which
are their own antiparticles $\gamma =\gamma ^{\dagger }$ \cite{Majorana}.
Although proposed originally as a model for neutrinos, MFs have recently
been predicted to occur as quasi-particle bound states in engineered
condensed matter systems \cite{AliceaJ}. This exotic particle obeys
non-Abelian statistics, which is one of important factors to realize
subsequent potential applications in decoherence-free quantum computation
\cite{BeenakkerCWJ,StanescuTD,FranzM} and quantum information processing
\cite{NayakC,ElliottSR}. Over the recent few years, the possibility for
hosting MFs in exotic solid state systems focused on topological
superconductors \cite{AliceaJ,BeenakkerCWJ,ElliottSR}. Currently, various
realistic platforms including topological insulators \cite{FuL1,FuL2},
semiconductor nanowires (SNWs) \cite{LutchynRM,OregY}, and atomic chains
\cite{Nadj-PergeS1,VazifehMM,KimY,PengY} have been proposed to support
Majorana states based on the superconducting proximity effect. Although
various schemes have been presented, observing the unique Majorana
signatures experimentally is still a challenging task to conquer.

MFs are their own antiparticles, and they are predicted to appear in
tunneling spectroscopy experiments, in which Majoranas manifest themselves
as characteristic zero-bias peaks (ZBPs) \cite{LawKT,FlensbergK1}. The
theoretical predictions of ZBPs have been observed experimentally in SNWs
which are interpreted as the signatures of MFs \cite%
{MourikV,DasA,DengMT,ChurchillHOH,FinckADK}. Remarkably, Nadj-Perge et al.
\cite{Nadj-PergeS2} recently designed a chain of magnetic Fe atoms deposited
on the surface of an s-wave superconducting Pb with strong spin-orbit
interactions, and reported the striking observation of a ZBP at the end of
the atomic chains\ with STM, which provides evidence for Majorana zero
modes. However, these above experimental results can not serve as definitive
evidences to prove the existence of MFs in condensed matter systems, and it
is also a major challenge in these experiments to uniquely distinguish
Majoranas from conventional fermionic subgap states. The first reason is
that the zero-bias conductance peaks can also appear in terms of the other
mechanisms \cite{LiuJ,ChangW}, such as the zero-bias anomaly due to Kondo
resonance \cite{FinckADK,LeeEJH} and the disorder or band bending in the SNW
\cite{BagretsD}. The second one is that Andreev bound states in a magnetic
field can also exhibit a zero-energy crossing as a function of exchange
interaction or Zeeman energy \cite{FrankeKJ,LeeEJH1}, and therefore give
rise to similar conductance features. As far as we know, most of the
experimental evidences for Majorana bound states largely relies on
measurements of the tunneling conductance at present, and the observation of
Majorana signature based on electrical methods still remains a subject of
debate. Identifying MFs only through tunnel spectroscopy is somewhat
problematic. Therefore, to obtain definitive signatures of MFs, alternative
setups or proposals for detecting MFs are necessary. Here, we will propose
an alternative all-optical scheme to detect MFs.

Benefitting from recent advances in nanotechnology and nanofabrication,
nanostructures such as quantum dots (QDs) and nanomechanical resonators
(NRs) have been obtained significant progress in modern nanoscience and
nanotechnology. QD, as a simple stationary atom with well optical property
\cite{JundtG}, lays the foundation for numerous possible applications \cite%
{UrbaszekB}. Due to high natural frequencies and large quality factors of
NRs \cite{PootM}, if QDs coupled to NRs \cite{Wilson-RaeI,YeoI,ChristineA}
to form hybrid systems, the coherent optical properties will be enhanced
remarkably, which will be an alternative ultrasensitive detection means.
Although probing MFs with QDs \cite{LiuDE,FlensbergK,LeijnseM,CaoYS,LiJ}
have been proposed, we notice that all the schemes are still based on
electrical means. In the present work, we propose an optical measurement
scheme to detect the existence of MFs in iron chains on the superconducting
Pb surface \cite{Nadj-PergeS2} via a coupled hybrid QD-NR system with
optical pump-probe scheme \cite{Xu}.

Compared with electrical detection means where the QDs are coupled to MFs
via the tunneling \cite{LiuDE,FlensbergK,LeijnseM,CaoYS,LiJ}, in our optical
scheme, there is no direct contact between MFs and the hybrid QD-NR system,
which can effectively avoid introducing\ other signals disturbing the
detecting of MFs. The interaction between MFs in iron chains and QD in
hybrid QD-NR system is mainly due to the dipole-dipole interaction, and the
distance between the two systems can be adjusted by several tens of
nanometers, therefore the tunneling between the QD and MFs can be neglected
safely. In addition, the QD is considered as a two-level system rather than
a single resonant level with spin-singlet state, and once MFs appear in\ the
end of iron chains and couple to the QD, the Majorana signature will be
carried out via the coherent optical spectrum of the QD. The change in the
coherent optical spectrum as a possible signature for MFs is another
potential evidence in the iron chains. This optical scheme will provide
another method for the detection of MFs, which is very different from the
zero-bias peak in the tunneling experiments \cite%
{MourikV,DasA,DengMT,ChurchillHOH,FinckADK,Nadj-PergeS2}. Furthermore, in
order to investigate the role of the NR in the hybrid system, we further
introduce the exciton resonance spectrum to detect MFs. The results shows
that the vibration of the NR acting as a phonon cavity will enhance the
exciton resonance spectrum significantly and make MFs more sensitive to be
detectable. The technique proposed here provide a new platform for
applications ranging from robust manipulation of MFs and MFs based quantum
information processing.

\section{MODEL AND THEORY}

Figure 1(b) shows the schematic setup that will be studied in this work,
where iron (Fe) chains on the superconducting Pb(110) surface \cite%
{Nadj-PergeS2}, and we employ a two-level QD with optical pump-probe
technology to detect MFs. The Fe chain is ferromagnetically ordered \cite%
{Nadj-PergeS2} with a large magnetic moment, which takes the role of the
magnetic field in the nanowire experiments \cite{MourikV}. Different from
the proposal of Mourik et al. \cite{MourikV}, this "magnetic field" is
mostly localized on the Fe chain, with small leakage outside, and
superconductivity is not destroyed along the chain. In this setup, the
energy scale of the exchange coupling of the Fe atoms is typically much
larger than that of the Rashba spin-orbit coupling and the superconducting
pairing. Figure 1(c) displays that a QD is implanted in the NR to form a
coupled hybrid QD-NR system. The whole system includes two kinds of
couplings which are QD-MF coupling and QD-NR coupling as shown in Fig. 1(a).
In the following, we will discuss the two kinds of coupling in detail,
respectively.

In the hybrid QD-NR system, the QD is modeled as a two-level system
consisting of the ground state $\left\vert g\right\rangle $ and the single
exciton state $\left\vert e\right\rangle $ at low temperatures \cite%
{ZrennerA,StuflerS}, and the Hamiltonian of the QD can be described as $%
H_{QD}=\hbar \omega _{e}S^{z}$ with the exciton frequency $\omega _{e}$,
where $S^{z}$ and $S^{\pm }$ are the pseudospin operator describing the
two-level exciton with the commutation relation $\left[ S^{z},S^{\pm }\right]
=\pm S^{\pm }$ and $\left[ S^{+},S^{-}\right] =2S^{z}$. For the NR, the
thickness of the beam is much smaller than its width, the lowest-energy
resonance corresponds to the fundamental flexural mode that will constitute
the resonator mode \cite{Wilson-RaeI} which can be characterized by a
quantum harmonic oscillator with Hamiltonian $H_{NR}=\hbar \omega
_{n}(b^{+}b+1/2)$, where $\omega _{n}$ is the resonator frequency and $b$ is
the annihilation operator of the resonator mode. Since the flexion induces
extensions and compressions in the structure \cite{Graff}, this longitudinal
strain will modify the energy of the electronic states of QD through
deformation potential coupling. Then the coupling between the resonator mode
and the QD is described by $H_{int}=\hbar \omega _{n}gS^{z}(b^{+}+b)$, where
$g$ is the coupling strength between the resonator mode and QD \cite%
{Wilson-RaeI}. Thus we obtain the Hamiltonian of the coupled hybrid QD-NR
system%
\begin{equation}
H_{QD-NR}=\hbar \omega _{e}S^{z}+\hbar \omega _{n}(b^{+}b+1/2)+\hbar \omega
_{n}gS^{z}(b^{+}+b).
\end{equation}

For the QD-MF coupling, as each MF is its own antiparticle, we introduce an
operator $\gamma $ with $\gamma ^{\dagger }=\gamma $ and $\gamma ^{2}=1$ to
describe MFs. Supposed that the QD couples to the nearby MF $\gamma _{1}$ in
the end of iron chains, then the Hamiltonian is written by \cite%
{LiuDE,FlensbergK,LeijnseM,CaoYS,LiJ}
\begin{equation}
H=i\epsilon _{M}\gamma _{1}\gamma _{2}/2+i\hbar \beta (S^{-}-S^{+})\gamma
_{1}.
\end{equation}%
To detect MFs, it is helpful to switch the Majorana representation to the
regular fermion one via the exact transformation $\gamma _{1}=f^{\dag }+f$
and $\gamma _{2}$ $=i(f^{\dag }-f)$, where $f$ and $f^{\dag }$ are the
fermion annihilation and creation operators obeying the anti-commutative
relation $\left\{ f\text{, }f^{\dag }\right\} =1$. Accordingly, in the
rotating wave approximation \cite{RidolfoA}, the above Hamiltonian can be
rewritten as%
\begin{equation}
H_{MF-QD}=\epsilon _{M}(f^{\dagger }f-1/2)+i\hbar \beta (S^{-}f^{\dag
}-S^{+}f),
\end{equation}%
where the first term gives the energy of MF with frequency $\omega _{M}$ and
$\epsilon _{M}=\hbar \omega _{M}\sim e^{-l/\xi }$ with the iron chains
length ($l$) and the Pb superconducting coherent length ($\xi $). If the
iron chains length ($l$) is large enough, $\epsilon _{M}$ will approach
zero. In the following, we will discuss the two conditions of $\epsilon
_{M}\neq 0$ and $\epsilon _{M}=0$, and define the two conditions as coupled
MFs ($\epsilon _{M}\neq 0$) and uncoupled MFs ($\epsilon _{M}=0$),
respectively. The second term describes the coupling between the nearby MF
and the QD with the coupling strength $\beta $, where the coupling strength
is related to the distance between the hybrid QD-NR system and the iron
chains. It should be also noted that the term of non-conservation for
energy, i.e. $i\hbar \beta (S^{-}f-S^{+}f^{+})$, is generally neglected. We
have made the numerical calculations (not shown in the following figures)
and shown that the effect of this term is too small to be considered in our
theoretical treatment.

Currently, the optical pump-probe technique has become a popular topic,
which affords an effective way to investigate the light-matter interaction.
The optical pump-probe technology includes a strong pump laser and a weak
probe laser \cite{Boyd}. In the optical pump-probe technology, the strong
pump laser is used to stimulate the system to generate coherent optical
effect, while the weak laser plays the role of probe laser. Therefore, the
linear and nonlinear optical effects can be observed via the probe
absorption spectrum based on the optical pump-probe scheme. Xu et al. have
obtained coherent optical spectroscopy of semiconductor QD when driven
simultaneously by two optical fields \cite{Xu}. Their results open the way
for the demonstration of numerous quantum level-based applications, such as
QD lasers, optical modulators, and quantum logic devices. In terms of this
scheme, we apply the pump-probe scheme to the QD of the hybrid QD-NR system
simultaneously. When the optical pump-probe technology is applied on the QD,
the Majorana signature will be carried out via the coherent optical
spectrum. The Hamiltonian of the exciton of the QD coupled to the two fields
is given by \cite{Boyd} $H_{P-QD}=-\mu
\sum\limits_{_{k=pu,pr}}E_{k}(S^{+}e^{-i\omega _{k}t}+S^{-}e^{i\omega _{k}t})
$, where $\mu $ is the dipole moment of the exciton, and $E_{k}$ is the
slowly varying envelope of the field.

Therefore, we obtain the whole Hamiltonian of the hybrid system as $%
H=H_{QD-NR}+H_{MF-QD}+H_{P-QD}$. In a rotating frame at the pump field
frequency $\omega _{pu}$, we obtain the total Hamiltonian of the system as
\begin{align}
H& =\hbar \Delta _{pu}S^{z}+\hbar \omega _{n}(b^{+}b+1/2)+\hbar \omega
_{n}gS^{z}(b^{+}+b)+\hbar \Delta _{M}(f^{\dag }f-1/2)+i\hbar \beta
(S^{-}f^{\dag }-S^{+}f)  \notag \\
& -\hbar \Omega _{pu}(S^{+}+S^{-})-\mu E_{pr}(S^{+}e^{-i\delta
t}+S^{-}e^{i\delta t}),
\end{align}%
where $\Delta _{pu}=\omega _{e}-\omega _{pu}$ is the detuning of the exciton
frequency and the pump frequency, $\Omega _{pu}=\mu E_{pu}/\hbar $ is the
Rabi frequency of the pump field, and $\delta =\omega _{pr}-\omega _{pu}$ is
the detuning of the probe field and the pump field. $\Delta _{M}=\omega
_{M}-\omega _{pu}$ is the detuning of the MF frequency and the pump
frequency. Actually, we have neglected the regular fermion like normal
electrons in the nanowire that interact with the QD in the above discussion.
To describe the interaction between the normal electrons and the exciton in
QD, a tight binding Hamiltonian of the whole iron chains is introduced \cite%
{ChenHJ}.

According to the Heisenberg equation of motion and introducing the
corresponding damping and noise terms, the quantum Langevin equations of the
whole system are derived as%
\begin{equation}
\dot{S}^{z}=-\Gamma _{1}(S^{z}+1/2)-\beta (S^{-}f^{+}+S^{+}f)+i\Omega
_{pu}(S^{+}-S^{-})+\dfrac{i\mu E_{pr}}{\hbar }(S^{+}e^{-i\delta
t}-S^{-}e^{i\delta t})\text{,}
\end{equation}%
\begin{equation}
\dot{S}^{-}=-[i(\Delta _{pu}+\omega _{n}gQ)+\Gamma _{2}]S^{-}+2(\beta
f-i\Omega _{pu})S^{z}-\frac{2i\mu E_{pr}}{\hbar }e^{-i\delta t}S^{z}+\hat{%
\tau}(t)\text{,}
\end{equation}%
\begin{equation}
\dot{f}=-(i\Delta _{M}+\kappa _{M}/2)f+\beta S^{-}++\hat{\varsigma}(t)\text{,%
}
\end{equation}%
\begin{equation}
\ddot{Q}+\gamma _{n}\dot{Q}+\omega _{n}^{2}Q=-2\omega _{n}^{2}gS^{z}+\hat{\xi%
}(t)\text{,}
\end{equation}%
where $\Gamma _{1}$ ($\Gamma _{2}$) is the exciton spontaneous emission rate
(dephasing rate), $Q=b^{+}+b$ is the position operator, $\gamma _{n}$ is the
decay rate of the NR, and $\kappa _{M}$ is the decay rate of the MF. $\hat{%
\tau}(t)$ is the $\delta $-correlated Langevin noise operator, which has
zero mean $\left\langle \hat{\tau}(t)\right\rangle =0$ and obeys the
correlation function $\left\langle \hat{\tau}(t)\hat{\tau}^{\dagger
}(t)\right\rangle \simeq \delta (t-t^{^{\prime }})$. The resonator mode is
affected by a Brownian stochastic force with zero mean value, and $\hat{\xi}%
(t)$ has the correlation function%
\begin{equation}
\left\langle \hat{\xi}^{+}(t)\hat{\xi}(t^{^{\prime }})\right\rangle =\dfrac{{%
\gamma }_{n}}{{\omega _{n}}}\int \dfrac{{d\omega }}{{2\pi }}\omega
e^{-i\omega (t-t^{^{\prime }})}[1+\coth ({\hbar \omega }/{2\kappa _{B}T})],
\end{equation}%
where $k_{B}$ and $T$ are the Boltzmann constant and the temperature of the
reservoir of the coupled system. MFs have the same correlation relation as
the resonator mode as
\begin{equation}
\left\langle \hat{\varsigma}^{+}(t)\hat{\varsigma}(t^{^{\prime
}})\right\rangle =\dfrac{{\kappa _{M}}}{{\omega _{M}}}\int \dfrac{{d\omega }%
}{{2\pi }}\omega e^{-i\omega (t-t^{^{\prime }})}[1+\coth ({\hbar \omega }/{%
2\kappa _{B}T})].
\end{equation}%
In Eq.(9) and Eq.(10), both the NR and Majorana mode will be affected by a
thermal bath of Brownian and non-Markovian processes \cite{GardinerCW}. In
the low temperature, the quantum effects of both the Majorana and NR mode
are only observed in the case of $\omega _{M}/{\kappa _{M}>>1}$ and ${\omega
_{n}}/{\gamma }_{n}{>>1}$. Due to the weak coupling to the thermal bath, the
Brownian noise operator can be modeled as Markovian processes. In addition,
both the QD-MFs coupling and QD-NR mode coupling in the hybrid system are
stronger than the coupling to the reservoir that influences the two kinds
coupling. In this case, owing to the second order approximation \cite%
{GardinerCW}, we can obtain the form of the reservoir that affects both the
NR mode and Majorana mode as Eq.(9) and Eq.(10).

To go beyond weak coupling, the Heisenberg operator can be rewritten as the
sum of its steady-state mean value and a small fluctuation with zero mean
value%
\begin{equation}
S^{z}=S_{0}^{z}+\delta S^{z},S^{-}=S_{0}+\delta S^{-},f=f_{0}+\delta
f,Q=Q_{0}+\delta Q
\end{equation}%
Since the driving fields are weak, but classical coherent fields, we will
identify all operators with their expectation values, and drop the quantum
and thermal noise terms. Simultaneously, inserting these operators into the
Langevin equations Eqs.(5)-(8) and neglecting the nonlinear term, we can
obtain two equation sets about the steady-state mean value and the small
fluctuation. The steady-state equation set consisting of $f_{0}$, $Q_{0}$
and $S_{0}$\ is related to the population inversion ($w_{0}=2S_{0}^{z}$) of
the exciton which is determined by%
\begin{gather}
\Gamma _{1}(w_{0}+1)[(\Delta _{M}^{2}+\kappa _{M}^{2}/4)(\Delta
_{pu}^{2}+\Gamma _{2}^{2}+\omega _{n}^{2}g^{4}w_{0}^{2}-2\omega _{n}\Delta
_{pu}g^{2}w_{0})  \notag \\
+\beta ^{2}w_{0}^{2}(\beta ^{2}-2\omega _{n}\Delta _{M}g^{2}+2\Delta
_{pu}\Delta _{M}-\Gamma _{2}\kappa _{M})]+4w_{0}\Gamma _{2}\Omega
_{pu}^{2}(\Delta _{M}^{2}+\kappa _{M}^{2}/4)=0.
\end{gather}%
For the equation set of small fluctuation, we make the ansatz \cite{Boyd} $%
\left\langle \delta O\right\rangle =O_{+}e^{-i\delta t}+O_{-}e^{i\delta t}$ (%
$O=S^{z},S^{-},f,Q$). Solving the equation set and working to the lowest
order in $E_{pr}$ but to all orders in $E_{pu}$, we can obtain the linear
susceptibility as $\chi _{eff}^{(1)}(\omega _{pr})=\mu S_{+}(\omega
_{pr})/E_{pr}=(\mu ^{2}/\hbar \Gamma _{2})\chi ^{(1)}(\omega _{pr})$, where $%
\chi ^{(1)}(\omega _{pr})$ is given by%
\begin{equation}
\chi ^{(1)}(\omega _{pr})=\frac{[{(\Pi _{4}^{\ast }+\Lambda }_{1}\Pi
_{3}^{\ast }{)\Pi _{1}\Lambda _{3}-iw_{0}\Pi _{4}^{\ast }}]\Gamma _{2}}{\Pi {%
_{2}\Pi _{4}^{\ast }-\Lambda _{1}\Lambda _{2}\Pi _{1}\Pi _{3}^{\ast }}},
\end{equation}%
$f_{0}$, $S_{0}$ and $Q_{0}$ can be derived from the steady-state equations,
and $\Sigma _{1}=\beta /(i\Delta _{M}+\kappa _{M}/2-i\delta )$, $\Sigma
_{2}=\beta /(-i\Delta _{M}+\kappa _{M}/2-i\delta )$, $\eta =2g\omega
_{n}^{2}/(\delta ^{2}+i\delta \gamma _{n}-\omega _{n}^{2})$, ${\Lambda _{1}}%
=[i\Omega _{pu}-\beta (f_{0}+S_{0}\Sigma _{2}^{\ast })]/(\Gamma _{1}-i\delta
)$, ${\Lambda _{2}}=[-i\Omega _{pu}-\beta (f_{0}^{\ast }+S_{0}^{\ast }\Sigma
_{1})]/(\Gamma _{1}-i\delta )$, ${\Lambda _{3}}=iS_{0}^{\ast }/(\Gamma
_{1}-i\delta )$, $\Pi _{1}=2(\beta f_{0}-i\Omega _{pu})-i\omega
_{n}gS_{0}\eta $, $\Pi _{2}=i(\Delta _{pu}-\delta +\omega _{n}gQ_{0})+\Gamma
_{2}-\beta w_{0}\Sigma _{1}-{\Lambda _{2}}\Pi _{1}$, $\Pi
_{3}=2(gf_{0}-i\Omega _{pu})-i\omega _{n}gS_{0}\eta ^{\ast }$, $\Pi
_{4}=i(\Delta _{pu}+\delta +\omega _{n}gQ_{0})+\Gamma _{2}-\beta w_{0}\Sigma
_{2}-{\Lambda _{3}}\Pi _{3}$ ($\Re ^{\ast }$ indicates the conjugate of $\Re
$). The imaginary and real parts of $\chi ^{(1)}(\omega _{pr})$ indicate
absorption and dissipation, respectively. In addition, the average
population of the exciton states can be obtained as%
\begin{equation}
S_{+}^{z}=\frac{({\Lambda _{1}\Pi _{3}^{\ast }+\Pi _{4}^{\ast }})[{\Lambda
_{3}(\Pi _{2}+\Lambda }_{2}\Pi _{1}{)-iw_{0}\Lambda _{2}}]}{\Pi {_{2}\Pi
_{4}^{\ast }-\Lambda _{1}\Lambda _{2}\Pi _{1}\Pi _{3}^{\ast }}},
\end{equation}%
which is benefited for readout the exciton states of QD.

\section{NUMERICAL RESULTS AND DISCUSSIONS}

For illustration of the numerical results, we choose the realistic hybrid
systems of the coupled QD-NR system \cite{Wilson-RaeI} and the iron chains
on the superconducting Pb surface \cite{Nadj-PergeS2}. For an InAs QD in the
coupled QD-NR system, we use parameters \cite{Wilson-RaeI}: the exciton
relaxation rate $\Gamma _{1}=0.3$ GHz, the exciton dephasing rate $\Gamma
_{2}=0.15$ GHz. The physical parameters of GaAs NR are $(\omega _{n}$, $M$, $%
Q_{f})=(1.2$ GHz, $5.3\times 10^{-18}$ kg, $3\times 10^{4})$, where $\omega
_{n}$, $M$, and $Q_{f}$ are the resonator frequency, the effective mass, and
quality factor of the NR, respectively. The decay rate of the NR is $\gamma
_{n}=$ $\omega _{n}/Q_{f}=40$ kHz, and the coupling strength between the QD
and NR is $g=0.06$. For MFs, there are no experimental values for the
lifetime of the MFs and the coupling strength between the exciton and MFs in
the recent literature. However, according to a few recent experimental
reports \cite{MourikV,DasA,DengMT,ChurchillHOH,FinckADK,Nadj-PergeS2}, it is
reasonable to assume that the lifetime of the MFs is $\kappa _{M}=0.1$ MHz.
Since the coupling strength between the QD and nearby MFs is dependent on
their distance, we also expect the coupling strength $\beta =0.05$ GHz via
adjusting the distance between the hybrid QD-NR system and the iron chains.

Figure 2(a) shows the coherent optical properties of the QD as functions of
probe-exciton detuning $\Delta _{pr}=\omega _{pr}-\omega _{e}$ at the
detuning of the exciton frequency and the pump frequency $\Delta _{pu}=0$,
i.e., the absorption ($Im\chi ^{(1)}$) and dissipation ($Re\chi ^{(1)}$)
properties of the QD without considering any coupling ($g=0,\beta =0$),
which indicates the normal absorption and dissipation of the QD,
respectively. Turning on the QD-NR coupling ($g=0.06$) and without
considering the QD-MF coupling ($\beta =0$), two sharp peaks will appear in
both the absorption and dissipation spectra as shown in Fig. 2(b). From the
curves, we find that the two sharp peaks at both sides of the spectra just
correspond to the vibrational frequency of the NR. The physical origin of
this result is due to mechanically induced coherent population oscillation,
which makes quantum interference between the resonator and the beat of the
two optical fields via the QD when the probe-pump detuning is equal to the
NR frequency \cite{LiJJ}. This reveals that if fixing the pump field
on-resonance with the exciton and scan through the frequency spectrum, the
two sharp peaks can obtain immediately in the coherent optical spectra,
which also indicates a scheme to measure the frequency of the NR. This
phenomenon stems from the quantum interference between the vibration NR and
the beat of the two optical fields via the exciton when probe-pump detuning $%
\delta $ is adjusted equal to the frequency of the NR. Therefore, the QD-NR
coupling play a key role in the hybrid system, and if we ignore the coupling
($g=0$), the above phenomenon will disappear completely as shown in Fig.
2(a).

Compared with Fig.2(b), in Fig.2(c), we consider the QD coupled with the
nearby MF $\gamma _{1}$ without taking the QD-NR coupling into account, i.e.
the condition of $g=0$ and $\beta =0.05$ GHz. As the MFs appear in the ends
of iron chains and coupled to the QD, both the probe absorption (the blue
curve) and dissipation (the green curve) spectra will present an remarkable
signature of MFs under $\Delta _{M}=-0.5$ GHz. The physical origin of this
result is due to the QD-MF coherent interaction and we can interpret this
physical phenomenon with dressed state between the exciton and MFs. The QD
coupled to the nearby MF will induce the upper level of the state $%
\left\vert e\right\rangle $ to split into $\left\vert e,n_{M}\right\rangle $
and $\left\vert e,n_{M}+1\right\rangle $ ($n_{M}$ denotes the number states
of the MFs). The left peak in the coherent optical spectra signifies the
transition from $\left\vert g\right\rangle $ to $\left\vert
e,n_{M}\right\rangle $ while the right peak is due to the transition of $%
\left\vert g\right\rangle $ to $\left\vert e,n_{M}+1\right\rangle $ \cite%
{ChenHJ}. To determine this signature is the true MFs rather than the normal
electrons that couple with the QD, we have used a tight binding Hamiltonian
to describe the electrons in whole iron chains, the numerical results
indicate the signals in the absorption and dissipation spectra are the true
MFs signature \cite{ChenHJ2}. If we consider both the two kinds coupling,
i.e. the QD-NR coupling ($g=0.06$) and QD-MFs coupling ($\beta =0.05$ GHz)
as shown in Fig. 2(d), not only the two sharp peaks locate at the NR
frequency induced by its vibration, i.e. two peaks are at $\Delta _{pr}=\pm
1.2$ GHz ($\omega _{n}=1.2$GHz), there is also MFs signal appear at $\Delta
_{pr}=-0.5$ GHz ($\Delta _{M}=-0.5$ GHz) induced by the QD-MF coupling.

In Fig. 2(c), we only consider the situation of $\epsilon _{M}\neq 0$. In
fact, if the iron chains length $l$ is much larger than the Pb
superconducting coherent length $\xi $, $\epsilon _{M}$ will approach zero.
Therefore, it is necessary to consider the conditions of $\epsilon _{M}\neq
0 $ and $\epsilon _{M}=0$, and we define them as coupled MFs mode ($\epsilon
_{M}\neq 0$) and uncoupled MFs mode ($\epsilon _{M}=0$), respectively.
Figure 3(a) and Figure 3(b) show the absorption and dissipation spectra as a
function of detuning $\Delta _{pr}$ with QD-MF coupling constants $\beta
=0.05$ GHz under $\epsilon _{M}\neq 0$ and $\epsilon _{M}=0$, respectively.
Compared with the coupled MFs mode, the uncoupled QD-MF Hamiltonian will
reduce to $H_{MF-QD}=i\hbar \beta (S^{-}f^{\dag }-S^{+}f)$ which is
analogous J-C Hamiltonian of standard model under $\epsilon _{M}=0$, and the
probe absorption spectrum (the blue curve) shows a symmetric splitting as
the QD-MF coupling strength $\beta =0.05$ GHz which is different from of
coupled MFs mode presenting unsymmetric splitting due to a detuning $\Delta
_{M}=-0.5$ GHz. Therefore, our results reveal that the signals in the
coherent optical spectra is a real signature of MF, and the optical
detection scheme can work at both the coupled Majorana edge states and the
uncoupled Majorana edge states.

In Fig. 3(c), we further make a comparison of the probe absorption spectrum
under the coupled MFs mode ($\epsilon _{M}\neq 0$) and uncoupled MFs mode ($%
\epsilon _{M}=0$). It is obvious that the probe absorption spectrum display
the analogous phenomenon of electromagnetically induced transparency (EIT)
\cite{FleischhauerM} under both the two conditions. The dip in the probe
absorption spectrum goes to zero at $\Delta _{pr}=0$ and $\Delta _{pr}=-0.5$
GHz with $\epsilon _{M}=0$ and $\epsilon _{M}\neq 0$, respectively, which
means the input probe field is transmitted to the coupled system without
absorption. Such a phenomenon is attributed to the destructive quantum
interference effect between the Majorana modes and the beat of the two
optical fields via the QD. If the beat frequency of two lasers $\delta $ is
close to the resonance frequency of MFs, the Majorana mode starts to
oscillate coherently, which results in Stokes-like ($\Delta _{S}=\omega
_{pu}-\omega _{M}$) and anti-Stokes-like ($\Delta _{AS}=\omega _{pu}+\omega
_{M}$) scattering of light from the QD. The Stokes-like scattering is
strongly suppressed because it is highly off-resonant with the exciton
frequency. However, the anti-Stokes-like field can interfere with the
near-resonant probe field and thus modify the probe field spectrum. Here the
Majorana modes play a vital role in this coupled system, and we can refer
the above phenomenon as Majorana modes induced transparency, which is
analogous with EIT in atomic systems \cite{FleischhauerM}.

On the other hand, we can propose a means to determine the QD-MF coupling
strength $\beta $ via measuring the distance of the two peaks with
increasing the QD-MF coupling strength in the probe absorption spectrum.
Figure 3(d) indicates the peak-splitting width as a function of the QD-MF
coupling strength $\beta $ under the condition of the coupled MFs mode ($%
\epsilon _{M}\neq 0$) and the uncoupled MFs mode ($\epsilon _{M}=0$) which
follows a nearly linear relationship. It is obvious that the two lines (the
uncoupled MFs and the coupled MFs mode) have a slight deviation. However,
the deviation becomes slighter with increasing coupling strength. Therefore,
it is essential to enhance the coupling strength for a clear peak splitting
via adjusting the distance between the QD and the nearby MFs. In this case
the coupling strength can obtain immediately by directly measuring the
distance of the two peaks in the probe absorption spectrum.

As shown in Fig. 2(d), there are not only two sharp peaks locate at the NR
frequency induced by its vibration but also the MFs signal appear at $\Delta
_{pr}=\Delta _{M}$ induced by the QD-MF coupling in the probe absorption
spectrum (the blue curve) under the two kinds coupling. In Fig. 4(a), we
further consider switching the detuning $\Delta _{M}=-0.5$ GHz to $\Delta
_{M}=-1.2$ GHz at small exciton-pump detuning $\Delta _{pu}=0.05$ GHz. It is
obvious that the resonance amplification process (1) and the resonance
absorption process (2) in the probe absorption spectrum without considering
the QD-MF coupling (the blue curve, $\beta =0$) will accordingly transform
into the the resonance absorption process (3) and the resonance
amplification process (4) due to the QD-MF coupling (the green curve, $\beta
=0.1$ GHz). Return to Fig. 1(a), there are two kinds of coupling which are
QD-NR coupling and QD-MF coupling in the hybrid system. For the QD-NR
system, the two sharp peaks in the probe absorption corresponding to the
resonance amplification (1) and absorption process (2)\ can be elaborated
with dressed states $\left\vert g,n\right\rangle $, $\left\vert
g,n+1\right\rangle $, $\left\vert e,n\right\rangle $, $\left\vert
e,n+1\right\rangle $ ($\left\vert n\right\rangle $ denotes the number state
of the resonance mode), and the two sharp peaks indicate the transition
between the dressed states \cite{LiJJ}. However, once MFs appear in the ends
of iron chains and coupled to the QD, the ground state $\left\vert
g\right\rangle $ and the exciton state $\left\vert e\right\rangle $ of the
QD will also modify by the number states of the MFs $n_{M}$ and induce the
Majorana dressed states $\left\vert g,n_{M}\right\rangle $, $\left\vert
g,n_{M}+1\right\rangle $, $\left\vert e,n_{M}\right\rangle $, $\left\vert
e,n_{M}+1\right\rangle $. With increasing the QD-MF coupling, the Majorana
dressed states will affect the amplification (1) and absorption process (2)
significantly, and even realize the inversion between the absorption (3) and
amplification (4) process due to the QD-MF coherent interaction (the green
curve).

To illustrate the advantage of the NR in the hybrid system, we introduce the
exciton resonance spectrum to investigate the role of NR in the coupled
QD-NR, which is benefited for readout the exciton states of QD. In Fig.
4(b), we adjust the detuning $\Delta _{M}=-0.5$ GHz to $\Delta _{M}=-1.2$
GHz, therefore, the location of the two sideband peaks induced by the QD-MF
coupling coincides with the two sharp peaks induced by the vibration of NR,
thus the NR is resonant with the coupled QD-MF system and makes the coherent
interaction of QD-MF more strong. Figure 4(b) shows the exciton resonance
spectrum of the probe field as a function of the probe detuning $\Delta
_{pr} $ with the detuning $\Delta _{pu}=0.05$ GHz under the coupled MFs mode
$\epsilon _{M}\neq 0$. The black and red curves correspond to $g=0$ and $%
g=0.06$ for the QD-MF coupling $\beta =0.1$ GHz, respectively. It is obvious
that the role of NR is to narrow and to increase the exciton resonance
spectrum. In this case, the NR behaves as a phonon cavity will enhance the
sensitivity for detecting MFs.

\section{CONCLUSION}

We have proposed an all-optical means to detect the existence of MFs in iron
chains on the superconducting Pb surface with a hybrid QD-NR system. The
signals in the coherent optical spectra indicate the possible Majorana
signature, which provides another supplement for detecting MFs. Due to the
vibration of NR, the exciton resonance spectrum becomes much more
significant and then enhances the detection sensitivity of MFs. In addition,
the QD-MF coupling in our system is a little feeble, while Ref. [35]
presents a strong QD-MF coupling and the coupling strength can reach
kilohertz, which is beneficial for MFs detection. On the other hand, if we
consider embedding a metal nanoparticle-quantum dot (MNP-QD) complex \cite%
{ChenHJ,LiJJ} in the NR, the surface plasmon induced by the MNP will enhance
the coherent optical property of QD, which may be robust for probing MFs.
The concept proposed here, based on the quantum interference between the NR
and the beat of the two optical fields, is the first all-optical means to
probe MFs. This coupled system will provide a platform for applications in
all-optically controlled topological quantum computing based on MFs.

\section{ACKNOWLEDGMENTS}

The authors gratefully acknowledge support from the National Natural Science
Foundation of China (No.11574206, No.10974133, No.11274230, No.61272153,
No.61572035, No.51502005, and No.11404005), the Key Foundation for Young
Talents in College of Anhui Province (NO. 2013SQRL026ZD), and the Foundation
for PhD in Anhui University of Science and Technology.

FIG.1 Sketch of the proposed setup for optically detecting Majorana fermions
(MFs). (a) The energy-level diagram of a QD coupled to MFs and NR, which
includes two kinds coupling, i.e. the QD-MF coupling (the dotted frame) and
the QD-NR coupling (the dashed frame). (b) The iron chains on the
superconducting Pb surface, and a pair of MFs appear in the ends of the iron
chains. The nearby MF is coupled to (c) a QD embedded in a nanomechanical
resonator (NR) with optical pump-probe technology.

FIG.2 The absorption (the blue curve) and dispersion (the green curve)
spectra of probe field as a function of the probe detuning $\Delta _{pr}$
under different conditions. (a) Without considering any coupling, i.e., $g=0$
and $\beta =0$. (b) The QD-NR coupling strength is $g=0.06$ and $\beta =0$.
(c) The QD-MF coupling strength is $\beta =0.05$ GHz and $g=0$. (d)
Considering both the QD-NR coupling and QD-MF coupling, i.e., $g=0.06$ and $%
\beta =0.05$ GHz. The parameters used are $\Gamma _{1}=0.3$ GHz, $\Gamma
_{2}=0.15$ GHz, $\gamma _{m}=40$ kHz, $\omega _{n}=1.2$ GHz, $\kappa
_{M}=0.1 $ MHz, $\Omega _{pu}^{2}=0.005$(GHz)$^{2}$, $\Delta _{M}=-0.5$ GHz,
and $\Delta _{pu}=0$.

FIG.3 (a) and (b) show the probe absorption (the blue curve) and dispersion
(the green curve) spectra with QD-MF coupling strengths $\beta =0.05$ GHz
under $\epsilon _{M}\neq 0$ and $\epsilon _{M}=0$, respectively. (c) The
probe absorption spectrum under $\epsilon _{M}\neq 0$ (the green curve) and $%
\epsilon _{M}=0$ (the blue curve), respectively. (d) The linear relationship
between the distance of peak splitting and the coupling strength of QD-MF $%
\beta $. The other parameters used are the same as in Fig.2.

FIG.4 (a) The probe absorption spectrum as a function of the probe detuning $%
\Delta _{pr}$ with considering (the blue curve, $\beta =0.1$ GHz) and
without considering (the green curve, $\beta =0$) the QD-MF coupling under
the QD-NR coupling strength $g=0.06$. (b) The exciton resonance spectrum as
a function of $\Delta _{pr}$ with $g=0$ and $g=0.06$ at the QD-MF coupling
strength $\beta =0.1$ GHz. $\Delta _{M}=-1.2$ GHz, $\Delta _{pu}=0.05$ GHz, $%
\Omega _{pu}^{2}=0.01$(GHz)$^{2}$, The other parameters used are the same as
Fig.2.

\clearpage
\begin{figure}[tbp]
\includegraphics[width=10cm]{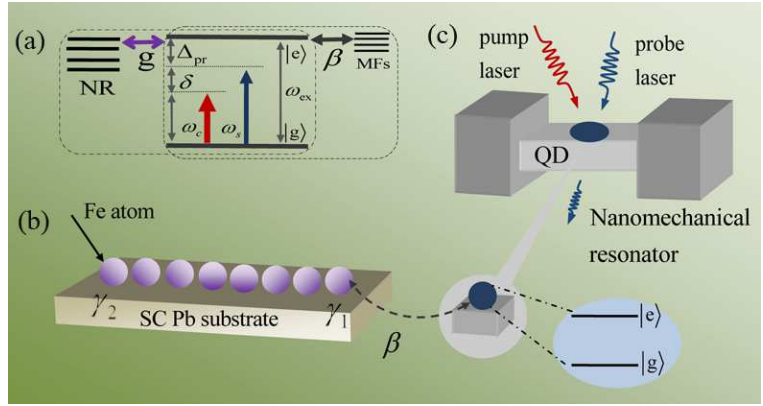}
\caption{Sketch of the proposed setup for optically detecting Majorana
fermions (MFs). (a) The energy-level diagram of a QD coupled to MFs and NR,
which includes two kinds coupling, i.e. the QD-MF coupling (the dotted
frame) and the QD-NR coupling (the dashed frame). (b) The iron chains on the
superconducting Pb surface, and a pair of MFs appear in the ends of the iron
chains. The nearby MF is coupled to (c) a QD embedded in a nanomechanical
resonator (NR) with optical pump-probe technology.}
\end{figure}

\begin{figure}[tbp]
\includegraphics[width=12cm]{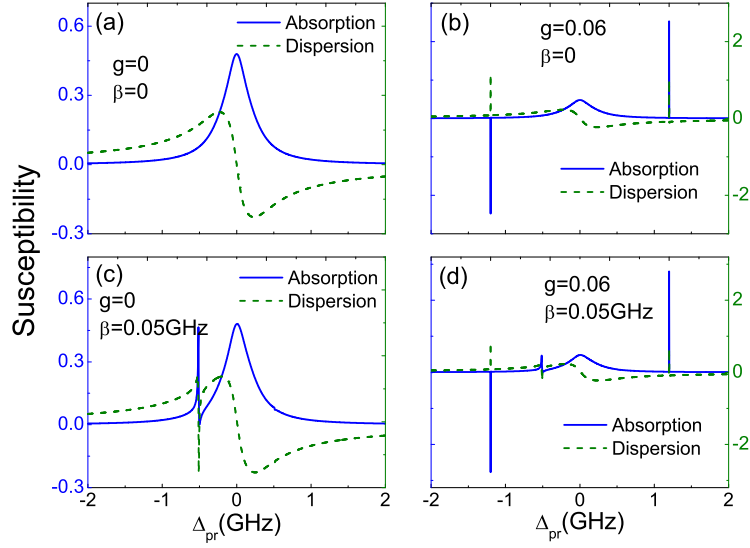}
\caption{The absorption (the blue curve) and dispersion (the green curve)
spectra of probe field as a function of the probe detuning $\Delta _{pr}$
under different conditions. (a) Without considering any coupling, i.e., $g=0$
and $\protect\beta =0$. (b) The QD-NR coupling strength is $g=0.06$ and $%
\protect\beta =0$. (c) The QD-MF coupling strength is $\protect\beta =0.05$
GHz and $g=0$. (d) Considering both the QD-NR coupling and QD-MF coupling,
i.e., $g=0.06$ and $\protect\beta =0.05$ GHz. The parameters used are $%
\Gamma _{1}=0.3$ GHz, $\Gamma _{2}=0.15$ GHz, $\protect\gamma _{m}=40$ kHz, $%
\protect\omega _{n}=1.2$ GHz, $\protect\kappa _{M}=0.1$ MHz, $\Omega
_{pu}^{2}=0.005$(GHz)$^{2}$, $\Delta _{M}=-0.5$ GHz, and $\Delta _{pu}=0$.}
\end{figure}

\begin{figure}[tbp]
\includegraphics[width=12cm]{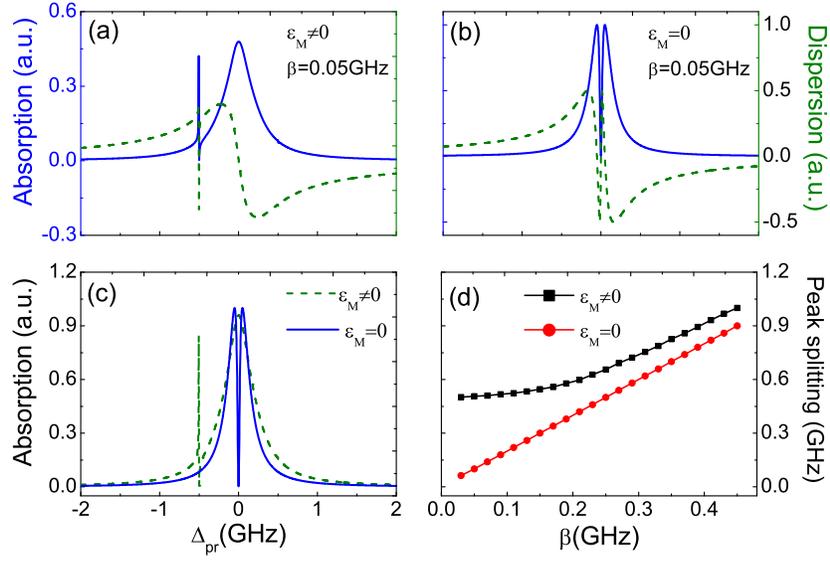}
\caption{(a) and (b) show the probe absorption (the blue curve) and
dispersion (the green curve) spectra with QD-MF coupling strengths $\protect%
\beta =0.05$ GHz under $\protect\epsilon _{M}\neq 0$ and $\protect\epsilon %
_{M}=0$, respectively. (c) The probe absorption spectrum under $\protect%
\epsilon _{M}\neq 0$ (the green curve) and $\protect\epsilon _{M}=0$ (the
blue curve), respectively. (d) The linear relationship between the distance
of peak splitting and the coupling strength of QD-MF $\protect\beta $. The
other parameters used are the same as in Fig.2.}
\end{figure}
\begin{figure}[tbp]
\includegraphics[width=12cm]{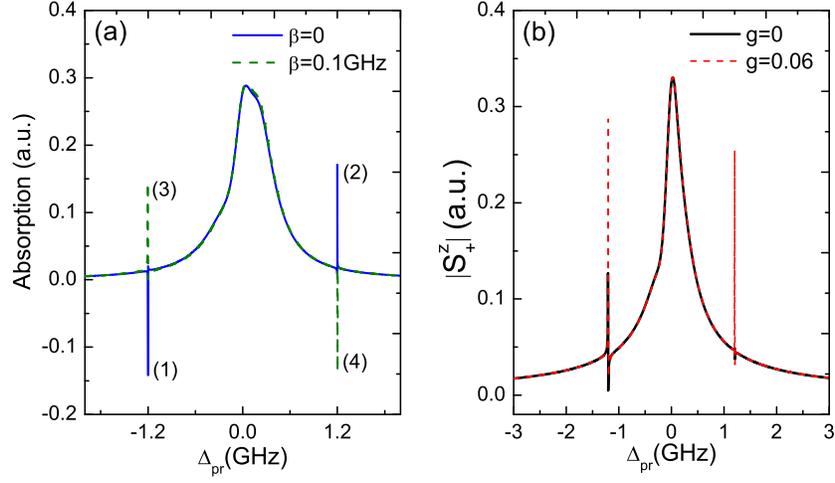}
\caption{(a) The probe absorption spectrum as a function of the probe
detuning $\Delta _{pr}$ with considering (the blue curve, $\protect\beta %
=0.1 $ GHz) and without considering (the green curve, $\protect\beta =0$)
the QD-MF coupling under the QD-NR coupling strength $g=0.06$. (b) The
exciton resonance spectrum as a function of $\Delta _{pr}$ with $g=0$ and $%
g=0.06$ at the QD-MF coupling strength $\protect\beta =0.1$ GHz. $\Delta
_{M}=-1.2$ GHz, $\Delta _{pu}=0.05$ GHz, $\Omega _{pu}^{2}=0.01$(GHz)$^{2}$,
The other parameters used are the same as Fig.2.}
\end{figure}

\end{document}